\documentstyle[12pt]{article}
\begin{document}
\renewcommand{\theequation}{\thesection.\arabic{equation}}
\newcommand{\eqn}[1]{(\ref{#1})}
\renewcommand{\section}[1]{\addtocounter{section}{1}
\vspace{5mm} \par \noindent
  {\large \bf \thesection . #1}\setcounter{subsection}{0}
  \par
   \vspace{2mm} } %was 5mm
\newcommand{\sectionsub}[1]{\addtocounter{section}{1}
\vspace{5mm} \par \noindent
  {\bf \thesection . #1}\setcounter{subsection}{0}\par}
\renewcommand{\subsection}[1]{\addtocounter{subsection}{1}
\vspace{2.5mm}\par\noindent {\bf  \thesubsection . #1}\par
 \vspace{0.5mm} }
\renewcommand{\thebibliography}[1]{ {\vspace{5mm}\par \noindent{\bf
References}\par \vspace{2mm}}
\list
 {\arabic{enumi}.}{\settowidth\labelwidth{[#1]}\leftmargin\labelwidth
 \advance\leftmargin\labelsep\addtolength{\topsep}{-4em}
 \usecounter{enumi}}
 \def\newblock{\hskip .11em plus .33em minus .07em}
 \sloppy\clubpenalty4000\widowpenalty4000
 \sfcode`\.=1000\relax \setlength{\itemsep}{-0.4em} }
\def\bea{\begin{eqnarray}}
\def\eea{\end{eqnarray}}
\def\be{\begin{equation}}
\def\ee{\end{equation}}
\def\alp{\alpha}
\def\bet{\beta}
\def\gam{\gamma}
\def\del{\delta}
\def\eps{\epsilon}
\def\sig{\sigma}
\def\lam{\lambda}
\def\Lam{\Lambda}
\def\m{\mu}
\def\n{\nu}
\def\r{\rho}
\def\s{\sigma}
\def\d{\delta}
\def\ss{\scriptstyle}
\newcommand\rf[1]{(\ref{#1})}
\def\nn{\nonumber}
\newcommand{\sect}[1]{\setcounter{equation}{0} \section{#1}}
\renewcommand{\theequation}{\thesection .\arabic{equation}}
\newcommand{\NPB}[3]{{Nucl.\ Phys.} {\bf B#1} (#2) #3}
\newcommand{\CMP}[3]{{Commun.\ Math.\ Phys.} {\bf #1} (#2) #3}
\newcommand{\PRD}[3]{{Phys.\ Rev.} {\bf D#1} (#2) #3}
\newcommand{\PLB}[3]{{Phys.\ Lett.} {\bf B#1} (#2) #3}
\newcommand{\JHEP}[3]{{JHEP} {\bf #1} (#2) #3}
\newcommand{\ft}[2]{{\textstyle\frac{#1}{#2}}\,}
\def\e{\epsilon}
\def\st{\scriptstyle}
\def\sst{\scriptscriptstyle}
\def\mco{\multicolumn}
\def\epp{\epsilon^{\prime}}
\def\vep{\varepsilon}
\def\ra{\rightarrow}
\def\ab{\bar{\alpha}}
\newcommand{\dt}{\partial_{\langle T\rangle}}
\newcommand{\dtbar}{\partial_{\langle\bar{T}\rangle}}
\newcommand{\al}{\alpha^{\prime}}
\newcommand{\mst}{M_{\scriptscriptstyle \!S}}
\newcommand{\mpl}{M_{\scriptscriptstyle \!P}}
\newcommand{\dv}{\int{\rm d}^4x\sqrt{g}}
\newcommand{\lv}{\left\langle}
\newcommand{\rv}{\right\rangle}
\newcommand{\ph}{\varphi}
\newcommand{\sbar}{\,\bar{\! S}}
\newcommand{\xbar}{\,\bar{\! X}}
\newcommand{\fbar}{\,\bar{\! F}}
\newcommand{\zbar}{\,\bar{\! Z}}
\newcommand{\tbar}{\bar{T}}
\newcommand{\ubar}{\bar{U}}
\newcommand{\ybar}{\bar{Y}}
\newcommand{\phb}{\bar{\varphi}}
\newcommand{\cm}{Commun.\ Math.\ Phys.~}
\newcommand{\pr}{Phys.\ Rev.\ D~}
\newcommand{\prl}{Phys.\ Rev.\ Lett.~}
\newcommand{\pl}{Phys.\ Lett.\ B~}
\newcommand{\ibar}{\bar{\imath}}
\newcommand{\jbar}{\bar{\jmath}}
\newcommand{\np}{Nucl.\ Phys.\ B~}
\newcommand{\gsi}{\,\raisebox{-0.13cm}{$\stackrel{\textstyle
>}{\textstyle\sim}$}\,}
\newcommand{\lsi}{\,\raisebox{-0.13cm}{$\stackrel{\textstyle
<}{\textstyle\sim}$}\,}

\thispagestyle{empty}

\begin{center}
\hfill AEI-092\\
\hfill UvA-WINS-Wisk-98-20\\
\hfill NIKHEF 98-029\\[3mm]
\hfill{\tt hep-th/9809070}\\

\vspace{2cm}

{\Large\bf D=11 SUGRA as the Low Energy Effective\\[3mm]
Action of  Matrix Theory: Three Form Scattering}\\
\vspace{1.4cm}
{\sc Jan Plefka ${}^{a,c}$, Marco Serone $^b$
and Andrew Waldron ${}^c$ }\\

\vspace{1.3cm}

${}^a${\em Albert-Einstein-Institut}\\
{\em Max-Planck-Institut f\"ur
Gravitationsphysik}\\
{\em Schlaatzweg 1, 14473 Potsdam} \\
{\em Germany}\\
{\footnotesize \tt plefka@aei-potsdam.mpg.de}\\

\vspace{.5cm}

${}^b$
{\em Department of Mathematics, University of Amsterdam }\\
{\em Plantage Muidergracht 24, 1018 TV Amsterdam} \\
{\em The Netherlands}\\
{\footnotesize\tt serone@wins.uva.nl}\\

\vspace{.5cm}

${}^c${\em NIKHEF, P.O. Box 41882, 1009 DB Amsterdam,}\\
{\em The Netherlands}\\
{\footnotesize \tt waldron@nikhef.nl}

\end{center}

\vspace{0.8cm}

\centerline{\bf Abstract}
\vspace{2 mm}  %\end{center}
\begin{quote}\small
We employ the LSZ reduction formula for Matrix Theory
introduced in our earlier work
to compute the $t$-pole $S$-matrix for three form--three form
scattering.
The result agrees completely with tree level
$D=11$ SUGRA. Taken together with previous results on
graviton-graviton scattering this shows that Matrix Theory
indeed reproduces the bosonic sector of the $D=11$ SUGRA
action including the Chern-Simons term.
Furthermore we provide a detailed account of our framework
along with the technology to compute any Matrix Theory
one-loop $t$-pole scattering amplitude at vanishing $p^-$
exchange.
\end{quote}
\vfill
\leftline{\sc September 1998}
\newpage
%%%%%%%%%%%%%%%%%%%%%%%%%%%%%%%%%%%%%%%%%%%%%%%%%%%%%%%%%%%%%%%%%
\setcounter{page}{1}
\baselineskip18pt

\addtocounter{section}{1}
\par \noindent
{\Large \bf \thesection . Introduction}
  \par
   \vspace{2mm} %was 5 mm
\noindent

%%%%%%%%%%%%%%%%%%%%%%%%%%%%%%%%%%%%%%%%%%%%%%%%%%%%%%%%%%%%%%%%%

It is now commonly believed that eleven dimensional supergravity
\cite{Cremmer}
is the low-energy effective theory of a more fundamental microscopic
theory, known as M-theory~\cite{witten}. A non-perturbative
definition of M-theory
has been conjectured to be given in terms of the large $N$ limit of  
a quantum
mechanical supersymmetric $U(N)$ Yang-Mills model called Matrix
Theory \cite{bfss}.
Since the time of this conjecture,
many computations in various  settings  have been performed to test the
proposal \footnote{See \cite{banks} for an exhaustive list of
references.}.
Most of these works, however, involve the comparison of classical
gravity
source-probe actions with the background field effective action of the
super Yang-Mills quantum mechanical system evaluated on straight line
configurations.
Clearly, however,  a principle test of the conjecture \cite{bfss}
would be to
compute the tree level $S$-matrix of $D=11$ SUGRA in Matrix
Theory.
We began this project in~\cite{psw} where we found, using a
formalism which enabled the computation of true scattering
amplitudes in Matrix Theory,
that indeed the
$D=11$ graviton--graviton tree level $t$-pole
S-matrix agrees precisely with that obtained
from Matrix Theory.
We stress that what was computed was
the full field theoretical amplitude, i.e. some 66 terms
depending on physical
polarisations and momenta, in contrast to previous works yielding
phase shifts in semi-classical eikonal scattering.

In this paper, we turn our attention away from the pure gravity sector of
the theory and consider three form scattering\footnote{The three  
form contribution to the linearized $D=11$ SUGRA potential has been
computed in Matrix Theory by~\cite{KT}.}
Again, making use of the leading effective potential for  
D-particles at one-loop~\cite{psw,mss1,mss2},
we find complete
agreement between the Matrix Theory and $D=11$ SUGRA $S$-matrices
(at tree level for the $t$-pole), an amplitude consisting of 103 terms.
Together with our previous results on graviton-graviton scattering  
\cite{psw}
this computation confirms that Matrix Theory describes all bosonic  
three-point interactions
of the $D=11$ supergravity action, including the Chern-Simons term
$\int dC\wedge dC\wedge C$. In this sense
$D=11$ SUGRA emerges as the low energy effective action of
Matrix Theory.

In addition we give a detailed account of the formalism
which could only be sketched in our previous letter~\cite{psw}, allowing
one to compute any $t$-pole zero $p^-$ exchange $S$-matrix element
in Matrix Theory at one loop.
Finally, as in \cite{psw}, throughout the paper we work in the
$N=2$ sector of the
matrix model, so that we are considering the
Susskind finite $N$ generalisation  \cite{suss} of the Matrix Theory
conjecture.

The paper is organised as follows.
The main idea of our framework is that $S$-matrix elements can be
constructed from
the asymptotic particle states of~\cite{pw} and involve the
expectation of
the usual Matrix Theory effective potential (including background
fermions)
between in and outgoing polarisation  states.
We refer to this relation between Matrix Theory effective actions  
and $S$-matrix elements as the Matrix Theory
Lehmann, Symanzik and Zimmermann (LSZ) reduction formula
and give a detailed derivation of this result in section two.
Although the one-loop Matrix Theory effective action could be
derived using quantum mechanical Feynman graphs it is obtained
more efficiently from the one-loop effective potential of two moving
D-particles, computed using the Green-Schwarz boundary state formalism
\cite{gregut1,mss1,mss2}. In section three we present a systematic
derivation of this result.
In section four
we combine the results of sections
two and three
to compute the Matrix Theory three form--three form scattering amplitude.
Furthermore we include a detailed account of the Matrix Theory kinematics
involved, along with
the algebra of spinor bilinear operators acting on
polarisation states, an essential ingredient
for efficient computations of such amplitudes.
Section five presents the $D=11$ SUGRA tree level computation
of the three form-three form scattering amplitude plus its
reduction to the vanishing $p^-$
exchange kinematics described by elastic $N=2$ Matrix Theory scattering.
Finally we give our conclusions and
in an appendix spell out some conventions and Fierz identities
needed for our computations.

\setcounter{equation}{0}
\section{LSZ for Matrix Theory}\label{LSZ}
\noindent
The purpose of this section is to provide the link between the
canonical operator based Matrix Theory scattering amplitude
calculations of \cite{pw} and the ``standard'' path integral,
background field, effective action approach
of, e.g., \cite{dkps,lm}. This provides the technology
required for scattering amplitude
calculations in Matrix Theory.

\subsection{Asymptotic Particle States in Matrix Theory}
\noindent
We begin with a short review of the asymptotic state analysis
of \cite{pw}. We shall primarily study the $U(2)$ Matrix Theory
Hamiltonian whose coordinates
take values in the adjoint representation
of $U(2)$ i.e.,
\bea
X_m &=& X^0_m \, i {\bf 1} + X^A_m \, i\sigma^A \qquad m=1,\ldots ,  
9 \nn\\
\theta &=&\theta^0 \, i {\bf 1} +\theta^A \, i\sigma^A
\eea
where $\sigma^A$ are the Pauli matrices. Employing a vector notation
for the $SU(2)$ part in which $\vec X_m=(X^1_m, X^2_m,X^3_m )$
and similarly for $\vec \theta$, the Matrix Theory Hamiltonian
is given
by\footnote{Note that we are using a real, symmetric representation of
the $SO(9)$ Dirac matrices in which the charge conjugation matrix
$C={\bf 1}$.}
\be
H= \ft 1 2 P^0_m P^0_m + \Bigl ( \ft 12 \vec{P}_m \cdot \vec{P}_m
+ \, \ft 14 (\vec{X}_m \times \vec{X}_n)^2
+ \, \ft i 2 \vec{X}_m\cdot \vec{\theta}\, \gamma_m\times
\vec{\theta}\Bigr )\, .
\label{MTHam}
\ee
It must be augmented by the Gauss law constraint
\be
\vec L= \vec X_m \times \vec P_m - \frac{i}{2}\vec \theta \times \vec
\theta
\label{GaussConstr}
\ee
which annihilates physical states. We wish to study particles
widely separated in (say) the ninth transverse direction, it is
therefore useful (but not necessary),
following~\cite{dWHN}, to
choose a frame
where (say) $\vec X_9$ lies along the $z$ axis
\be
X^1_9=0=X^2_9 \, . \label{frame}
\ee
Denoting the Cartan coordinates $x_m=X^3_m$ and the remainder
$Y^I_a=X^I_a$ (with $I=1,2$ and $a=1,\ldots,8$)
one then sees that the Hamiltonian \eqn{MTHam}
takes the form
\be
H= H_{\rm CoM}(X^0) + H_V(x_m) + H_{\rm HO}(Y^I_a,\theta^I|x_m)+
H_4(Y^I_a,x_m,\theta^I,\theta^3)
\label{MTHamGF}
\ee
Here $H_{\rm CoM}=-\ft 12 (\partial_{X^0_m})^2 $ is the $U(1)$
center of mass Hamiltonian. In particular, one now observes that
$H_V=-\ft {1}{2x_9} (\partial_{x_m})^2 x_9$ represents free particle
propagation
along the ``Cartan valley'', whereas $H_{\rm HO}$
describes a system of 16 superharmonic oscillators transverse to
the Cartan valley with frequency $r=(x_m x_m)^{1/2}$
depending on the distance from  the valley origin. Finally
$H_4$ constitutes all remaining terms.

In \cite{pw} it was shown that in the limit $x_9\rightarrow \infty$,
which one interprets as the large separation in the nine direction
between a pair of
asymptotic particles (importantly, observe that in a general frame,
$x_9=(\vec{X}_9\cdot\vec{X}_9)^{1/2}$ is gauge invariant,
i.e. commutes with the constraint), there exists a split of the
Hamiltonian into a free and an interacting part (the latter
of which is suppressed in the large
$x_9$ limit). The free Hamiltonian admits eigenstates of the form
\be
|p^1_m,{\cal H}^1;p^2_m,{\cal H}^2\rangle=|0_B,0_F\rangle_{x_m}\,
\ft{1}{x_9}e^{i(p^1-p^2) \cdot
x}e^{i(p^1+p^2)\cdot X^0}
|{\cal H}^1\rangle_{\theta^0+\theta^3}\,|{\cal H}^2
\rangle_{\theta^0-\theta^3}\, .
\label{state}
\ee
Here $|0_B,0_F\rangle_{x_m}$ denotes the superharmonic
groundstate of $H_{\rm HO}$
with vanishing zero point energy. Note that the oscillator states depend
explicitly on the Cartan coordinates $x_m$ through their frequency
$r$.
In the above, $p^I_m$ are the momenta of the two particles and  
their possible polarisations ${\cal H}^I$ are those of the graviton,
three-form tensor and gravitino represented by states
\be
|h\rangle = h_{mn} |-\rangle^{mn}, \qquad
|C\rangle = C_{mnp} |-\rangle^{mnp}, \qquad
|\psi\rangle = \psi_m^\alpha |-\rangle_\alpha^m \, ,
\ee
whose explicit form was constructed in \cite{pw}. Finally, the subscripts
$\theta^0\pm\theta^3$ in \eqn{state} indicate from which
fermionic variables the polarisation states are built. The state
\eqn{state} is
free in the asymptotic limit, i.e.,
\be
\lim_{x_9\rightarrow \infty} H |p^1_m,{\cal H}^1;p^2_m,{\cal H}^2\rangle
= \ft 1 2 \left [ (p^1)^2 + (p^2)^2\right ]
|p^1_m,{\cal H}^1;p^2_m,{\cal H}^2\rangle
\ee
for large particle separations $x_9$. Moreover, the state \eqn{state}
indeed satisfies the physical state condition
$\vec L \, |p^1_m,{\cal H}^1;p^2_m,{\cal H}^2\rangle=0$.

\subsection{The LSZ Reduction Formula for Matrix Theory}

We now turn to the computation of scattering amplitudes and  
derive,in particular, the formula relating Matrix Theory effective  
action
computations with the scattering matrix. For a
$1 + 2 \rightarrow 4+3$ process one starts with the $S$-matrix
element
\bea
S_{fi}&=&
\lefteqn{\langle p^4,{\cal H}^4;p^3,{\cal H}^3|\exp \{-iHT\} |
p^1,{\cal H}^1;p^2,{\cal H}^2\rangle} \nonumber \\
&=&
\int d^9 X'^{0}\, 4\pi {x}_9'{}^2 d^9 x' \,\,
 d^9 X^0\,\, 4\pi x_9{}^2 d^9 x\,
\ft{1}{x'_9}e^{-i(p^4-p^3)_m x'_m}
e^{-i(p_4+p_3)_m X'^0_m}
\nonumber\\&&
{}_{_{_{_{\theta^0+\theta^3}}}}\!\!\!\langle {\cal H}_4|\,
{}_{_{_{_{\theta^0-\theta^3}}}}\!\!\!\langle {\cal H}_3|
\,\,{}_{x'_m}\!
\langle 0_B,0_F|\,\exp(-iHT)\,|0_B,0_F\rangle
\!{}_{x_m}
\,\, |{\cal H}_1\rangle\!\!_{_{_{_{\theta^0+\theta^3}}}}\,
|{\cal H}_2\rangle\!\!_{_{_{_{\theta^0-\theta^3}}}}\nonumber\\
&& \quad\quad \ft{1}{x_9}e^{i(p^1-p^2)_m x_m} e^{i(p^1+p^2)_m X^0_m}
\label{S}
\eea
where $T$ is the large time during which the process takes place.
The measure factors $4\pi {x_9}^2$ appear in the integral
because of
the choice of frame
\eqn{frame}.
Moreover, since we are interested in eikonal kinematics, we have
chosen asymptotic states describing particles widely separated in  
the nine direction for both the in and out states. More general
configurations
may be handled via the insertion of a rotation operator in~\eqn{S}.

For complete clarity, we note the following. Strictly speaking,
one should
compute $S_{fi}=\langle {\rm out}|\exp(-iHT)| {\rm in}\rangle$
with $H$ being the  Hamiltonian \eqn{MTHam} in a general frame
and the asymptotic states $| {\rm in}\rangle$ and $\langle {\rm out}|$
written in a manifestly gauge invariant way {\it without} fixing  
the frame
$X^I_9=0$, as shown in~\cite{pw}. Now as
$\vec{L}|{\rm in}\rangle=0=\langle {\rm out}|\vec{L}\,\,$
and $[H,\vec{L}]=0$ we have
\be
S_{fi}=\langle {\rm out}|\exp(-iHT)| {\rm in}\rangle=
\langle {\rm out}|\Pi\exp(-iHT)\Pi| {\rm in}\rangle
\ee
where $\Pi=({\rm vol}_{SU(2)})^{-1}\int
d\vec{\Omega}_{SU(2)}\,\exp(i\vec{\Omega}\cdot \vec{L})$
is the projector onto gauge invariant states. Therefore
one is able to choose a frame $X'{}^I_9=0=X^I_9$ at {\it both} start and
endpoints and in this way arrives at \eqn{S} in which
the variables of the fixed frame~\eqn{frame} appear.

The vacuum to vacuum transition amplitude
${}_{x'_m}\!
\langle 0_B,0_F|\,\exp(-iHT)\,|0_B,0_F\rangle
\!{}_{x_m}$ is now the object of interest and
may be represented as a path integral with appropriate boundary
values for the Cartan coordinates
\bea
\lefteqn{e^{i\Gamma(x_m,x'_m,\theta^3)-iH_{\rm CoM}}
\equiv
{}_{x'_m}\!
\langle 0_B,0_F|\,\exp(-iHT)\,|0_B,0_F\rangle
\!_{x_m}=}\nn\\&&
{\cal N} \int_{x_m(-T/2)=x_m,\, \theta^3(-T/2)=\theta^3}^{x_m(T/2)=
x'_m,\, \theta^3(T/2)=\theta^3}
D^{16}Y D^{16}\theta^I D^9x_m \, (4\pi {x_9}^2) \exp ( i\int_{-T/2}^{T/2}
dt L)
\label{PI}
\eea
Here the Lagrangian $L$ is the usual one (``$p\dot{q}-H$'')
associated with the Hamiltonian
\eqn{MTHamGF} in the special frame ($X^I_9=0$). In particular, as a  
result
of this choice of frame we have a measure
factor $4\pi{x_9}^2$ at each point in the path which may be
exponentiated
via ghosts. ${\cal N}$ denotes some normalisation
factor\footnote{In fact,
the observant reader will note that in what follows, we assume that  
the normalisation factor ${\cal N}$ behaves as $(x'_9\, x_9)^{-1}$  
to
cancel the
measure
factors of the initial and final integrations over the valley
coordinates
$x_m$. That this is the case can be argued by $SO(9)$ covariance of  
the final
result. We also do not compute the overall normalisation of the
path integral.
Such technicalities should, in principle, be rigorously calculable via
a careful construction of the path integral similar to that presented
in~\cite{de Boer}.}. Importantly, note that the boundary conditions  
for the
oscillator variables $Y^I_a$ and $\theta^I$ are zero at start and
endpoints
since we are considering a transition between harmonic oscillator ground
states in these variables.

It is also essential to observe that the transition
element appearing in \eqn{S} depends on
{\it operator} valued $\theta^3$, whereas in \eqn{PI} one computes  
a $c$-number
valued path integral. To elevate this result to an operator, as
needed in the
rest of the computation, one makes only errors proportional to the
square of the momentum transfer $q^2$ which we anyway neglect in this
work since they correspond to contact terms not detectable in our
$D0$-brane
computation. To see that the $\theta^3$ boundary conditions in the
transition amplitude \eqn{PI}
are correct, one can change from the sixteen real variables $\theta^3$
to eight complex ones and perform a coherent state analysis similar to
that of~\cite{Peeters}.

We now make the following observation. Firstly, consider the
BRST gauge fixed path integral of the ten dimensional super Yang-Mills
dimensionally reduced to quantum mechanics
\be
e^{i\Gamma(x_m,x'_m,\theta^3)}={\cal N}\!\int_{x_m(-T/2)=x_m,\,
\theta^3(-T/2)=\theta^3}^{x_m(T/2)=
x'_m,\, \theta^3(T/2)=\theta^3}
D\vec A \, D^9\vec X_m \, D^{16}\vec\theta \, D\vec b \,
D\vec c \,\exp ( i\int_{-T/2}^{T/2}
dt \, L_{\rm BRST})
\label{PIBRS}
\ee
where $\vec b$ and $\vec c$ are $SU(2)$ ghosts,
$L_{\rm BRS}(\vec A,\vec X,\vec \theta,\vec b,\vec c)$
is the dimensionally reduced super Yang-Mills
Lagrangian with ghost terms and the gauge field
$\vec A$
is the time component of the ten dimensional Yang-Mills field.
This is the path integral considered in most Matrix Theory
computations (including the boundary conditions quoted in \eqn{PIBRS}).
Then, if one takes the gauge choice $A^3=0=X^I_9$ and additionally  
integrates
out the ghosts and remaining gauge field components $A^I$
(yielding, respectively, the measure factor $4\pi x^2$ and the frame
fixed Lagrangian), one obtains {\it exactly} the path integral
\eqn{PI}. The ramifications of this simple observation are clear;
one may now obtain $S$-matrix elements from the effective actions
produced
by the usual Matrix Theory computations found in the
literature~\cite{banks}.
The path integral~\eqn{PIBRS} can be computed by expanding about
classical
trajectories $X_m^3\equiv x_m^{\rm cl}(t)=b_m+v_m t$ and constant
$\theta^3(t)=\theta^3$ which satisfy the quoted boundary conditions
for impact parameter $b_m=(x_m'+x_m)/2$ and velocity
$v_m=(x_m'-x_m)/T$. A quantum mechanical Feynman diagram expansion
in the gauge of one's choice then leads to the required effective action
although we found it more efficient to exploit the connection
between the Matrix Theory and String Theory D0-brane dynamics
in order to obtain $\Gamma(x_m',x_m,\theta^3)$ (see section three). 
Our LSZ reduction formula for Matrix Theory, relating the effective  
action to
$S$-matrix elements is therefore
\bea
S_{fi}&=&\int d^9 X'^{0}\, d^9 x' \,\, d^9 X^0\,d^9 x\nonumber\\
&&
\exp\Big(-i(p^4-p^3)_m x_m'-i(p_4+p_3)_m X'^0_m
+i(p^1-p^2)_m x_m+i(p^1+p^2)_m X^0_m\Big)\nonumber\\
&&\quad\quad\quad\quad
{}_{_{_{_{\theta^0+\theta^3}}}}\!\!\!\langle {\cal H}_4|\,
{}_{_{_{_{\theta^0-\theta^3}}}}\!\!\!\langle {\cal H}_3|
e^{i\Gamma(x_m,x'_m,\theta^3)-iH_{CoM}T}
|{\cal H}_1\rangle\!\!_{_{_{_{\theta^0+\theta^3}}}}\,
|{\cal H}_2\rangle\!\!_{_{_{_{\theta^0-\theta^3}}}}\,\,.
\label{reduce}
\eea
Finally, we complete this section by noting that the generalisation  
of this
formula to higher $N$ Matrix Theory, i.e., $SU(N)$, $N$ particle
elastic scattering with vanishing $p^-$ momentum exchange is
straightforward.
One needs only replace the pairs of incoming and outgoing
polarisation states in~\eqn{reduce} by a set of $N$ such states.
The effective
action becomes that of an $SU(N)$ Matrix Theory computation depending on
$N-1$ Cartan coordinates and the momentum plane waves become those  
of $N$ in
and outgoing particles.

%%%%%%%%%%%%%%%%%%%%%%%%%%%%%%%%%%%%%%%%%%%%%%%%%%%%%%%%%%%%%%%%%

\setcounter{equation}{0}
\section{String Computation of the $D0$-Brane Effective Potential}
\label{string}
\noindent
In this section, after a very brief review of the Green-Schwarz
boundary
state formalism \cite{gregut1}, we
give a detailed account of the
computation of the one-loop
Matrix Theory potential (see equation (8) of \cite{psw}),
first performed in \cite{mss2,psw}. We consider here the
D-particle case, but it is clear from
\cite{mss1,mss2} that our result is trivially extendable to generic  
p-branes.

Dp-brane defects \cite{pol} can be described in String Theory
by suitable boundary states
implementing
the usual Neumann-Dirichlet boundary conditions, both in the  
covariant\cite{pocai,clny,li} as well as the Green-Schwarz formalism
\cite{gregut1}.
In the latter framework, the boundary state
describing
a single flat
D-brane is defined
by the BPS condition
\be
Q^a_+|B\rangle = 0  \;,\; Q^{\dot a}_+ |B\rangle = 0  \, ,\label{bps}
\ee
where $Q^a_+,Q^{\dot{a}}_+$ are suitable linear combinations of the  
$SO(8)$
supercharges
$Q^a,\widetilde{Q}^a,Q^{\dot{a}},\widetilde{Q}^{\dot{a}}$. The
solution for
$|B\rangle$ turns out to be
\be
\label{mom}
|B\rangle =\exp\sum_{n>0}\left({1\over n}  M_{ij}
\alpha^i_{-n}\widetilde{\alpha}^j_{-n} -
i M_{a\dot{a}}S^a_{-n}\widetilde{S}^{\dot{a}}_{-n}\right)|B_0\rangle
\ee
$|B_0\rangle$ being the zero mode part
\be
\label{zm}
|B_0\rangle= M_{ij} |i\rangle|\widetilde j\rangle
- i M_{\dot a \dot b}|\dot a\rangle|\widetilde{\dot b}\rangle
\label{zero}
\ee
with $M_{ij},M_{a\dot{a}},M_{\dot{a}a}$ definite $SO(8)$ matrices
\cite{gregut1,mss1},
depending on the dimensionality of the brane \footnote{In writing
$M_{a\dot{a}}$ we have
implicitly chosen to work in the IIA theory, relevant for the
analysis of D-particles.}.
In this gauge, the $\pm$ light-cone directions satisfy
automatically Dirichlet
boundary conditions, meaning that they are, effectively, Euclidean  
branes.
One might think that
defining the boundary state for moving branes by simply
boosting the static
one \cite{billo} would then be problematical,
however, as explained in \cite{mss1,mss2}, it is
possible to overcome this
difficulty by identifying one of the $SO(8)$ transverse directions with
the
time direction. Thereafter one deduces
the corresponding $SO(1,9)$ expressions and performs a double
analytic continuation
to the final covariant result.

Given these preliminaries, one may then compute arbitrary one-loop  
n-point functions of vertex operators $V_1,\ldots,V_n$
\be
{\cal A}_n=\int_0^\infty \!\!dt \,
\langle B,\vec{x}|e^{-2\pi t\al p^+(P^--i\partial/\partial
x^+)}V_1\dots V_n
| B,\vec{y} \rangle \label{cyli}
\ee
with $P^-=\frac{(p^i)^2}{p^+}+{\rm osc.}$ the light-cone  
Hamiltonian(the term $i\partial/\partial x^+$ just implements
the $p^-$ subtraction needed to obtain the effective Hamiltonian in  
this gauge). The
configuration space boundary state $|B,\vec{x}\rangle$ is given by
\be
\label{conf}
|B,\vec{x}\rangle=(2\pi\sqrt{\al})^{4-p}\int\frac{d^{9-p}q}{(2\pi)^{9-p}}
\,e^{i \vec{q}\cdot\vec{x}}\,|B\rangle\otimes|\vec{q}\rangle
\ee \noindent
with $\langle q|q^{\prime}\rangle
={\rm vol}_{p+1}\,(2\pi)^{9-p}\delta^{(9-p)}(q-q^{\prime})$
and ${\rm vol}_{p+1}$ is the space-time volume spanned by the p-brane.
Equation (\ref{cyli}) describes the interaction of D-branes,  
considered as semiclassical
heavy objects, with $n$ arbitrary states, described by the vertex
operators $V_i$.
We are interested in computing the leading one-loop
effective action
$\Gamma(v,\theta_3,r)$ of two moving D-particles with relative velocity
$v$ at their minimum separation $r$, that is, the usual
$v^4 / r^7$ term plus
all other terms, bilinear in fermions, related to it by
supersymmetry. Correspondingly,
we will consider the following correlator, encoding in particular
the abovementioned terms
\be
{\cal V}=\frac{1}{2}\int_0^\infty \!\!dt \,
\langle B,\vec{x}=0|e^{-2\pi t\alpha^{\prime}
p^+(P^--i\partial/\partial x^+)}
e^{(\eta
Q_-+\widetilde{\eta}\widetilde{Q}_-)}e^{V_B}|B,\vec{y}=\vec{b}
\rangle
\label{corr}
\ee
$Q_-,\widetilde{Q}_-$ being the SO($8$) supercharges broken by the  
presence
of D-branes and $V_B$ the vertex operator
that boosts the branes to a relative velocity $v_i$, given explicity by
\be
V_B=v_i\oint_{\tau=0}\!d\sigma\left(X^{[1}\partial_{\sigma}X^{i]}
+\frac{1}{2}S\,\gamma^{1i}S\right) \label{boost}
\ee
where the direction 1 will be Wick rotated to give the time
direction.
The factor 1/2 in (\ref{corr}) has been introduced
in order to normalise the
$v^4$ term to one.

A configuration of parallel branes preserves 1/2 of the
supercharges; in light-cone
gauge this implies that among the 16 linearly realised supercharges
$S^a_0,\widetilde{S}_0^{\dot{a}}$, eight of them are left unbroken.
Equations (\ref{cyli}) and (\ref{corr}) require  then the insertion  
of at least eight
zero modes
(that, due to the constraint (\ref{bps}), can be always chosen to  
be $S_0^a$)
in order to get a non-vanishing result. This is precisely the
number of zero modes provided
by the $v^4/r^7$ term and all its related fermionic terms, which  
are therefore
determined by
massless string excitations alone \cite{mss1,mss2}. In this configuration
we can then consider
only the massless part of (\ref{corr}) which simplifies  
dramatically. Integrating over the cylinder modulus $t$, we obtain
\be
{\cal V}=T\, T_0^2 \!\int\!
\frac{d^{9}q}{(2\pi)^{9}}\frac{e^{i \vec{q}\cdot\vec{b}}}{q^2}
\langle B_0|e^{(\eta
Q^0_-+\widetilde{\eta}\widetilde{Q}^0_-)}e^{V_B^{(F)}}|B_0\rangle
\label{corr0}
\ee
where $Q^0_-,\widetilde{Q}^0_-$ are just the zero mode part of the
given supercharges, $V_B^{(F)}$ is the fermionic part of the boost  
operator
(\ref{boost}), $T_0=\sqrt{\pi}(4\pi^2\al)^{3/2}$ is the tension of  
0-branes,
T is the overall time in which the interaction takes place and
$\vec{q}$ spans the directions $\pm$,2,...,8. The bosonic
part of $V_B$ induces conservation of momentum that reads
$q^1=\vec{q}\cdot\vec{v}$;
we fix in (\ref{corr0}) and in the following
$q^1=\vec{q}\cdot\vec{v}=0$, which simply
means we are computing the effective potential at the
time\footnote{From now on for simplicity
we will omit the overall time of the interaction $T$ and
the 0-brane tension $T_0$.} $t=0$.
It is now convenient to write the $S_0$ zero mode trace in terms  
of$R_0^{ij}=(S_0^a\gamma^{ij}_{ab}S_0^b)/4$.
Expanding the exponentials we find, following \cite{mss2},
\bea
{\cal V}_m&=&\!\sum_{n,m:\atop n/2+m=4}\!\frac{2^m}{n!m!}
\int\frac{d^{9}q}{(2\pi)^{9}}\frac{e^{i \vec{q}\cdot\vec{b}}}{q^2}
t^{1i_1...1i_mj_1j_2...j_{n-1}j_{n}} \nonumber \\
&\!\!\!\!\!\!& v_{i_1}...v_{i_m}
(\sqrt{q^+}\eta+\frac{\widetilde{\eta}q_l\gamma^l}{\sqrt{q^+}})^{[a_1}
...
(\sqrt{q^+}\eta+\frac{\widetilde{\eta}q_m\gamma^m}{\sqrt{q^+}})^{a_n]}
\gamma^{j_1 j_2}_{a_1a_2} ... \gamma^{j_{n-1} j_{n}}_{a_{n-1}a_n}
\label{corr0s}
\eea
where $t^{i_1...i_8}$ is the usual eight-tensor
\be
t^{i_1...i_8} \equiv
\mbox{Tr}_{S_0}\,R_0^{i_1i_2}R_0^{i_3i_4}R_0^{i_5i_6}R_0^{i_7i_8}
\ee
whose explicit form can be found, e.g., in the appendix of
chapter nine, volume II of \cite{gsw}.
Although the explicit computation for the $v^4,v^3$ and $v^2$ cases  
has been already
performed in \cite{mss2}, we will report them here for completeness

{\bf $v^4$-term}:
\be
{\cal V}_4=\frac{2^4}{4!}\int\!\frac{d^{9}q}{(2\pi)^{9}}
\frac{e^{i\vec{q}\cdot\vec{b}}}{q^2}t^{1i1j1k1l}v_iv_jv_kv_l
=G_{9}(\vec{b})\, v^4
\ee
where $G_d(\vec{b})$ is the propagator for a scalar massless  
particle in d-dimensions.

{\bf $v^3$-term}:
\bea
\hspace{-1.0cm} &\!&{\cal V}_3=\frac{2^3}{2!3!}\!
\int\!\frac{d^{9}q}{(2\pi)^{9}}\frac{e^{i \vec{q}\cdot\vec{b}}}{q^2}
t^{1i1j1klm}v_iv_jv_k
(q^+\eta\gamma^{lm}\eta+2\,q_n\eta\gamma^{lm}\gamma^n\widetilde{\eta}
+\frac{q_nq_p}
{q^+}\widetilde{\eta}\gamma^{n}\gamma^{lm}\gamma^p\widetilde{\eta})  
\nonumber \\
&\,&=2\,v^2 v_m
\int\!\frac{d^{9}q}{(2\pi)^{9}}\frac{e^{i \vec{q}\cdot\vec{b}}}{q^2}
(q^+\eta\gamma^{1m}\eta+2\,q_n\eta\gamma^{1mn}\widetilde{\eta}
+\frac{q^2_{l.c.}}{q^+}
\widetilde{\eta}\gamma^{1m}\widetilde{\eta}) \label{t3}
\eea
where $q^2_{l.c.}=\sum_{i=2}^{8}q_iq_i$ and $q^{\pm}=q^0\pm q^9$;
notice that
$q^2=q^2_{l.c.}-q^+q^-$ implying
that $q^-=q^2_{l.c.}/q^+$, modulo contact terms that are vanishing  
in our configuration of
separated 0-branes. It is trivial to verify that the
$SO(1,9)$ expression for the term in parenthesis in (\ref{t3}) is
$\bar\psi\Gamma^{1mn}\psi\,q_n$ (our conventions
for the Dirac matrices are given in the appendix),
with $n=\pm,2,...8$, which
after analytic continuation (that sends
also $v^i\rightarrow iv^i$) leads to
\be
{\cal V}_3=-2i\,v^2 v_m J^{0mn} \,\partial_n G_{9}(\vec{b})=
2i\,v^2 v_m (\theta\gamma^{mn}\theta)\,\partial_n G_{9}(\vec{b})
\ee
where $J^{\mu\n\rho}\equiv\bar\Psi\Gamma^{\mu\n\rho}\Psi$, $\Psi$ is
the Majorana-Weyl spinor $\Psi=\left(^{\textstyle
\theta}_{\textstyle 0}\right)$,
$\theta = (\eta^a, \tilde \eta^{\dot a})$ and $m,n=1,...,9$.
We have written the result also
in terms of the $SO(9)$ spinor $\theta$ which  is  the
most convenient way to connect
it to Matrix Theory.
It is clear that we can from now on restrict our attention to  
terms with no net power of
$q^+$; as shown in (\ref{t3}), these terms contain enough
information to reconstruct
the covariant form of our amplitudes..

{\bf $v^2$-term}:
\be
{\cal V}_2=\frac{2^2\cdot 6}{2!4!}
\int\frac{d^{9}q}{(2\pi)^{9}}\frac{e^{i \vec{q}\cdot\vec{b}}}{q^2}
t^{1i1jlmnp}\,v_iv_j\,\omega_{lmnp}(q,\eta)
\ee
where we defined
\be
\omega_{i_1...i_{2n}}(\eta,q) \equiv \frac{1}{(2n!)} \eta_{[a_1}
(\tilde \eta \hat{q})_{a_2}
... \eta_{a_{2n-1}} (\tilde \eta \hat{q})_{a_{2n}]}
\gamma^{i_1 i_2}_{a_1a_2} ... \gamma^{i_{2n-1} i_{2n}}_{a_{2n-1}
a_{2n}}\label{anti}
\ee
and $\hat{q}=q_i\gamma^i$. It is not difficult to see that
\be
t^{1i1jlmnp}v_iv_j\omega_{lmnp}(q,\eta)=
\frac{2}{3}v^2(J^{1mq}J^{1n}_{\;\;\;\;\; q}+
J^{mi\mu}J^{\;\;nj}_{\mu}\hat{v}_i\hat{v}_j)q_m\,q_n
\label{v2c}
\ee
with latin indices labelling the indices $i,j,...=\pm,2,...8$,
whereas greek indices
run over all 10 directions. Strictly speaking the equality
(\ref{v2c}) holds
only for $m,n=2,...,8$. The cases where $m,n=\pm$ are given by the  
terms in (\ref{corr0s})
with non-vanishing powers of $q^+$, as can be easily checked. This  
is the sense in which
(\ref{v2c}) and similar identities that will follow should be
understood.
We can now analytically continue the right hand side of
(\ref{v2c}) leading to
\be
{\cal V}_2=\frac{1}{3}v^2
( J^{m 0q} J^{n}_{\;\;\; 0q} + J^{m \mu}_{\;\;\;\;\; i}
J^{n}_{\;\; \mu j} \hat v^i \hat v^j)
\partial_m \partial_n G_{9}(\vec{b})=-2(\theta\gamma^{pm}\theta)
(\theta\gamma^{qn}\theta)
v_p\,v_q \,\partial_m \partial_n G_{9}(\vec{b})
\ee
where the $SO(9)$ expression follows after a Fierz identity.

{\bf $v^1$-term}:
\be
{\cal V}_1=\frac{2\cdot 20}{6!}
\int\!\frac{d^{9}q}{(2\pi)^{9}}\frac{e^{i \vec{q}\cdot\vec{b}}}{q^2}
t^{1ij_1...j_6}\,v_i\, \omega_{j_1...j_6}(q,\eta)
\ee
This case, as well as the next one, is a little more involved,
since we have a new
contribution
\be
t^{1ij_1...j_6}v_i
\omega_{j_1...j_6}(q,\eta)=12\,\omega_{1ijkkj}(q,\eta)+
24\,\omega_{1jjkki}(q,\eta)-1/2\,\epsilon^{1ij_1...j_6}
\omega_{j_1...j_6}(q,\eta)
\label{1v1}
\ee
The first term on the right hand side of (\ref{1v1}) is
vanishing due to the Fierz
identity (\ref{fi4}). The second term is
\be
\omega_{1jjkki}(q,\eta)=\frac{1}{20}J^{1nl}J_{l\mu}^{\; \;\;
m}J^{\mu\,i\,p} q_n q_m q_p
\label{2v1}
\ee
Using the relations (\ref{fi7}) reported in the appendix, it
is possible to verify that the $SO(1,9)$ expression for
the $\epsilon$-term is, up to a numerical factor
\be
\epsilon_{1\mu_1 ...\mu_9}\,q^{\mu_9}\,v^{\mu_8}\,q_{\alpha}\,q_{\beta}\,
J^{\mu_1\mu_2\alpha}J^{\mu_3\mu_4\beta}J^{\mu_5\mu_6\mu_7}
\label{3v1}
\ee
The expression (\ref{3v1}) can be brought into the same form as
the right hand side of
(\ref{2v1}) using
the identity (\ref{fi5}). In order to fix the relative
coefficient
between the two non-vanishing contributions coming from (\ref{1v1}),
it is much simpler to consider the term
proportional to $(q^+)^3$,
in which case the spinor algebra simplifies considerably. This term  
is proportional to
\be
-1/2\,\epsilon^{1ij_1...j_6}(\eta\gamma^{j_1j_2}\eta)(\eta\gamma^{j_3j_4}\eta)
(\eta\gamma^{j_5j_6}\eta)+24(\eta\gamma^{1j}\eta)(\eta\gamma^{jk}\eta)
(\eta\gamma^{ki}\eta)
\label{4v1}
\ee
By using the identity (\ref{fi6}), the $\epsilon$-term in
(\ref{4v1}) becomes
$8(\eta\gamma^{1j}\eta)(\eta\gamma^{jk}\eta)(\eta\gamma^{ki}\eta)$.  
Putting all the
results together we find
\be
{\cal V}_1=\frac{4i}{45}v_i\,J^{0nl}J_{l\mu}^{\; \;\; m}J^{\mu\, i\, p}
\,\partial_m \partial_n \partial_p G_{9}(\vec{b})
=-\frac{4i}{9} v_i (\theta\gamma^{im}\theta) (\theta\gamma^{nl}\theta)
(\theta\gamma^{pl}\theta)\,\partial_m \partial_n \partial_p
G_{9}(\vec{b})
\label{V1}
\ee
where the second identity in (\ref{V1}) follows from the first
one by $SO(9)$
Fierz identities.

{\bf $v^0$-term}:
\be
{\cal V}_0=\frac{70}{8!}\int\frac{d^{9}q}{(2\pi)^{9}}\frac{e^{i
\vec{q}\cdot\vec{b}}}{q^2}
t^{i_1...i_8} \omega_{i_1...j_8}(q,\eta)
\ee
where
\be
t^{i_1...i_8} \omega_{i_1...i_8}(q,\eta)=24\,\omega_{ijjkklli}(q,\eta)
-1/2\,\epsilon^{i_1...i_8}\omega_{i_1...i_8}(q,\eta)
\label{1v0}
\ee
The $SO(1,9)$ expression for the $\epsilon$-term is
\be
\epsilon_{\mu_1 ...\mu_{10}}q^{\mu_{10}}q_{\alpha}q_{\beta}q_{\gamma}
J^{\mu_1\mu_2\alpha}J^{\mu_3\mu_4\beta}J^{\mu_5\mu_6\gamma}
J^{\mu_7\mu_8\mu_9}
\label{2v0}
\ee
whereas
\be
\omega_{ijjkklli}(q,\eta)=\frac{1}{70}J^{\mu\nu m}J_{\mu}^{\; \;\rho n}
J_{\nu}^{\;\;\sigma p}J_{\rho\sigma}^{\; \;\;\; q}\,q_m\, q_n\, q_p\, q_q
\label{3v0}
\ee
Again, the expression (\ref{2v0}) can be cast in the form appearing  
on the right hand side
of (\ref{3v0}). By looking at the $(q^+)^4$-term, similarly to
the previous case,
it turns out that the $\epsilon$-term in (\ref{1v0}) gives a
contribution equal to
$8\,\omega_{ijjkklli}(q,\eta)$. We then obtain
\bea
{\cal V}_0&=&\frac{32}{8!} J^{\mu\nu m}J_{\mu}^{\; \;\rho n}
J_{\nu}^{\;\;\sigma p}J_{\rho\sigma}^{\; \; \;\;q}\,
\partial_m \partial_n \partial_p \partial_q G_{9}(\vec{b}) \nonumber \\
&=&\frac{2}{63}(\theta\gamma^{ml}\theta) (\theta\gamma^{nl}\theta)
(\theta\gamma^{pk}\theta)(\theta\gamma^{qk}\theta)
\,\partial_m \partial_n \partial_p \partial_q G_{9}(\vec{b})
\label{V0}
\eea
where once again $SO(9)$ Fierz identities have been used to write
the second identity in (\ref{V0}).

Collecting all terms we obtain the final result for the leading
one-loop potential of two D0-branes \cite{psw}
\bea
{\cal V}^{(1)}&=&\Bigl [\, v^4 + 2i\,v^2 v_m
(\theta\gamma^{mn}\theta)\,\partial_n
-2v_p\,v_q (\theta\gamma^{pm}\theta) (\theta\gamma^{qn}\theta)
\,\partial_m \partial_n \nonumber \\
&&\quad - \frac{4i}{9} v_i (\theta\gamma^{im}\theta)
(\theta\gamma^{nl}\theta)
(\theta\gamma^{pl}\theta)\,\partial_m \partial_n \partial_p
\label{POT} \\
&&\quad + \frac{2}{63}(\theta\gamma^{ml}\theta) (\theta\gamma^{nl}\theta)
(\theta\gamma^{pk}\theta)(\theta\gamma^{qk}\theta)
\,\partial_m \partial_n \partial_p \partial_q \Bigl ] \,
G_{9}(\vec{b}) \nonumber
\eea
The first, second, third and last terms of (\ref{POT}) were
calculated in a
super Yang-Mills context in \cite{dkps},\cite{Kraus97},\cite{arth}  
and\cite{static} respectively.
The supersymmetry parameter $\theta$
should be
identified with the spinor $\theta^3/2$ introduced in
the previous section and represents
the fermionic background in Matrix Theory.

Before concluding the present section we want to make some comments  
about the origin of the
$\epsilon$-terms in (\ref{1v1}),(\ref{1v0}). By performing an
analysis of 1-point
functions of massless closed string states on a disc with
supercharges inserted on
its boundary, it is straightforward to derive which fields
exchanged between the branes
are responsible for the interactions described above
\cite{gregut2,mss1,mss2}.
In this way, as one might expect, all the interactions, except
those coming from the
$\epsilon$-terms, are due to exchange of dilatons, gravitons and
Ramond-Ramond (RR)
vector gauge fields. On the
other hand, the $\epsilon$-terms arise from an interesting coupling  
between dual
RR gauge potentials, very similar to that analysed in  
\cite{bis,D8,D6}.In particular the
$\epsilon$-term coming from the part of the potential linear in the  
velocity is
due to exchange of a RR one form $A_{(1)}$ and its dual seven-form  
$A_{(7)}$.
The insertion of six supercharges corresponding to D-particles
on the boundary of the disc
indeed induces a non-minimal coupling to the seven-form.
Schematically, in light-cone gauge
it reads
\be
\langle B|Q^6|A_{(7)}\rangle\sim \omega_{j_1...j_6}(q) A_{[1j_1...j_6]}
\label{eps1}
\ee
where the direction 1 satisfies the Neumann boundary condition on  
the disc and
will be identified with the time direction. The coupling  
(\ref{eps1}) produces then an interaction
\be
\langle B|Q^6|A_{(7)}\rangle \langle A_{(7)}|A_{(1)}\rangle \langle  
A_{(1)}|V_B|B\rangle
\sim \frac{1}{q^2}v_i\, \omega_{j_1...j_6}(q)\,\epsilon^{1ij_1...j_6}
\ee
where we used the relation between dual transverse gauge potentials  
$A_{(1)}=\,^*\!\!A_{(7)}$.
Notice the relationship of this interaction to the corresponding
one analysed in
\cite{D6} for the D0-D6 brane system. \\
The $\epsilon$-term associated to the static potential can be
treated similarly.
Again, the RR seven-form $A_{(7)}$ has a non-vanishing 1-point function
when eight supercharges are inserted, that includes a term
\be
\langle B|Q^8|A_{(7)}\rangle\sim \omega_{1j_1...j_7}(q) \,
A_{[j_1...j_7]}
\label{eps2}
\ee
leading to an interaction
\be
\langle B|Q^8|A_{(7)}\rangle \langle A_{(7)}|A_{(1)}\rangle \langle  
A_{(1)}|B\rangle
\sim \frac{1}{q^2}  \, \omega_{j_1...j_8}(q) \,
\epsilon^{j_1...j_8}
\ee
equal to the $\epsilon$-term appearing in the static potential.
In this case the correspondence with the analogous static RR
potential for the D0-D8
system, found in \cite{D8}, is more subtle. Indeed, it has been
shown in \cite{D8}
that non-physical polarisations of the RR nine-form, identified
with some RR one-form
polarisations, are responsible for the RR attraction between a D0
and a D8 brane.
Although these effects are clearly not visible in a transverse
physical gauge, it
can be seen in a covariant formalism that the RR nine-form has a
non-vanishing 1-point
function with D0-branes, when eight supercharges are inserted.
The considerations above are of course only schematic but the key  
point is to highlight the presence of such interesting interactions  
which can be analysed in more detail along the lines of  
\cite{bis,D8,D6}.

\setcounter{equation}{0}
\section{Three Form Scattering in Matrix Theory}
\label{matrixcomp}
\noindent

In this section we present the results for three form scattering
in Matrix Theory. The section will be divided into three parts. In  
the first,
we spell out carefully the kinematics of the scattering amplitude under
consideration. In the second we develop
the algebra of bilinears built from the
$SO(9)$ fermionic operators $\theta^3$ acting on polarisation states.
This latter development allows one to perform the Matrix Theory
computation
of $S$-matrix elements, once one is given the potential as in~\eqn{POT},
with comparable (or even improved) efficiency to that when employing
tree level Feynman diagrams in $D=11$ supergravity.
In the third section we state our result.

\subsection{Kinematics}

The starting point is our Matrix Theory LSZ formula~\eqn{reduce} which
yields the $S$-matrix for the $1+2\longrightarrow 4+3$
scattering of particles with momenta $p_m^1$, $p_m^2$, $p_m^4$,
and $p_m^3$, respectively.
To begin with, the free $U(1)$ center of mass
sector of the theory decouples and yields an overall factor,
$
(2\pi)^9\delta^9(P_m'-P_m)\exp(-iP_m P_m T/2)
$,
expressing
conservation of total momentum. Here $P_m=p^1_m+p^2_m$
and $P_m'=p_m^4+p_m^3$ are the
total in and outgoing momenta, respectively, and the exponential
is the standard factor obtained in
time independent perturbation theory\footnote{In what follows, we shall
disregard these kinematical prefactors in comparing to
the SUGRA Feynman graph result since they only express the usual
relation between time independent and time dependent perturbation
theory.}.

A loopwise expansion of the Matrix Theory effective action yields
\be
\Gamma(x_m',x_m,\theta^3)=
\Gamma(b_m,v_m,\theta^3)=
v_m v_m T/2 +\Gamma^{({\rm loops})}
\ee
with $v_m=(x_m'-x_m)/T$ and $b_m=(x_m'+x_m)/2$, so that
the $S$ matrix now reads (up to an overall normalisation)
\be
S_{fi}=\int d^9x'd^9x\exp(-iw_m x_m'+iu_m x_m+iv_m v_m T/2)
\langle {\cal H}^3|\,\langle {\cal H}^4|\,
e^{i\Gamma^{({\rm loops})}(v_m,b_m,\theta^3)}\,
|{\cal H}^1\rangle\,|{\cal H}^2\rangle
\ee
where we denoted the in and outgoing relative momenta
$u_m=(p^1_m-p_m^2)/2$ and $w_m=(p^4_m-p_m^3)/2$, respectively.
However, changing variables $d^9x'd^9x\rightarrow d^9(Tv)d^9b$,
the integral over $Tv_m$ may be performed, for large $T$, by stationary
phase which yields
\be
S_{fi}=e^{-i[(u+w)/2]^2}\,\int d^9b \, e^{-iq_m b_m}
\langle {\cal H}^3|\,\langle {\cal H}^4|\,
e^{i\Gamma^{({\rm loops})}(v_m=(u+w)_m/2,b_m,\theta^3)}\,
|{\cal H}^1\rangle\,|{\cal H}^2\rangle\label{nearly}
\ee
where $\Gamma^{({\rm loops})}$ is to be evaluated at $v_m=(u+w)_m/2$,
which we take, henceforth, as the definition of $v_m$.
Moreover $q_m$ denotes the momentum transfer $q_m=w_m-u_m$.
For clarity, let us give the relation between the various momenta
introduced
above
\bea
p^1_m=P_m+u_m=P_m+v_m-q_m/2 &&
p^2_m=P_m-u_m=P_m-v_m+q_m/2 \nonumber\\
p^4_m=P_m'+w_m=P_m+v_m+q_m/2 &&
p^3_m=P_m'-w_m=P_m-v_m-q_m/2 \nonumber
\eea
Energy-momentum conservation fixes, in particular, the important
relation $q_m v_m=0$.

Finally, we make one last manipulation. The Matrix Theory effective  
action
is an expression of the form $\Gamma^{({\rm
loops})}=\int_{-T/2}^{T/2}dt\,
{\cal V}^{({\rm loops})}[x_m^{\rm cl}=(b_m+v_m t),\dot x_m^{\rm cl}
=v_m,\theta^3]$
for some potential ${\cal V}^{({\rm loops})}$.
However,
we would really like to consider a time independent potential depending
on $b_m$ and $v^m$ and $\theta^3$ acting on states. This is achieved
by expanding the exponential inside the polarisation expectation
value in~\eqn{nearly} and then interchanging the $dt$ and $d^9b$
integrations.
Then shifting $b_m\rightarrow b_m+v_m t$ and using $q_m v_m=0$,
the $dt$ integral yields only an  overall factor $T$ (the same factor
$T$ as appearing in front of equation~\eqn{corr0} in our string
computation).
Finally, we have the desired expression for the one-loop Matrix Theory
$S$-matrix
\be
S_{fi}=iT\exp(-iv^2)\int\! d^9b \, e^{-iq_m b_m}
\langle {\cal H}^3|\,\langle {\cal H}^4|\,
{\cal V}^{(1)}(b_m, v_m,\theta^3)
|{\cal H}^1\rangle\,|{\cal H}^2\rangle\, .
\label{lightatendoftunnel}
\ee
The loopwise expansion of the effective action is valid for large  
impact parameters $b_m$ and hence small momentum transfer $q_m$.
In section five we will show that this is precisely the limit
dominated by $t$-channel tree level physics in the SUGRA computation.

\subsection{Algebra of Bilinears}

All that remains to complete our Matrix Theory
computation is to perform a Fourier transform with respect to
$b_m$ of~\eqn{lightatendoftunnel} and insert the one-loop Matrix Theory
potential~\eqn{POT} into the polarisation inner products.
The result is an amplitude which may be directly compared with the
result of the SUGRA tree level Feynman diagram calculation of
section five.

In our previous work, when computing the graviton-graviton scattering
amplitude we used the explicit representation for the polarisation states
in terms of $SO(7)\otimes U(1)$ covariant complex spinors
(see~\cite{pw}).
Then rewriting the Matrix Theory potential also in terms of complex
spinors, we were able to compute its  polarisation expectation value.
Unfortunately, if one considers now the three form polarisation
states, this naive method becomes rather quickly unwieldy.
Instead, we now present an $SO(9)$ covariant algebra of spinor bilinears
acting on polarisation states with which amplitudes of complicated Matrix
Theory potentials can be efficiently computed.

The Matrix Theory potential (\ref{POT}) is written solely in terms  
of $SO(9)$ rotation generators $\ft18\theta^3\gamma^{mn}\theta^3$  
where the operators $\theta^3_{\alpha}$ satisfy the anticommutation  
relations
$\{\theta^3_\alpha,\theta^3_\beta\}=\delta_{\alpha\beta}$.
However, the states
\be
|{\cal H}^{1,4}\rangle=
{\cal H}^{1,4}_{{\cal M}}\,|-\rangle_{\theta^0+\theta^3}^{{\cal
M}}\, ,\qquad
|{\cal H}^{2,3}\rangle={\cal H}^{2,3}_{{\cal
M}}\,|-\rangle_{\theta^0-\theta^3}
^{{\cal M}}\label{M}
\ee
(where the generalised index ${\cal M}$ denotes
${\cal M}\equiv \{mn\, ;\, mnp\, ;m\alpha\}$, corresponding to
graviton, three-form and gravitino polarisations, respectively)
are built from states depending either on the sum or difference
of the centre of mass
($\{\theta^0_\alpha,\theta^0_\beta\}=\delta_{\alpha\beta}$)
and Cartan fermionic coordinates
which we denote as
$\theta^1=(\theta^0+\theta^3)/\sqrt{2}$ and
$\theta^2=(\theta^0-\theta^3)/\sqrt{2}$.
Of course, the operators $\frac{1}{8}\theta^1\gamma^{mn}\theta^1$
and $\frac{1}{8}\theta^2\gamma^{mn}\theta^2$ act simply as $SO(9)$  
rotation
generators on the states $|-\rangle^{\cal M}_{\theta^1}$ and
$|-\rangle^{\cal M}_{\theta^2}$, respectively,
under which they transform in the usual fashion. Therefore, if we
could write
the potential only in terms of these operators the computation  
would be completely trivial. Yet there are  cross terms since
\be
\frac{1}{8}\theta^3\gamma^{mn}\theta^3
=
\frac{1}{4}\theta^1\gamma^{mn}\theta^1+
\frac{1}{4}\theta^2\gamma^{mn}\theta^2-
\frac{1}{2}\theta^1\gamma^{mn}\theta^2\,
\ee
Let us now concentrate on amplitudes involving bose particles only
(the generalisation to the fermi case is simple but not needed here).
Clearly, the difficult terms are those involving the bilinear
$\theta^1\gamma^{mn}\theta^2$. However, since between bose states
the expectation of an odd number of $\theta$ operators vanishes
(and the 1 and 2 sectors are independent), only an even number of  
such operators can occur. But products $\theta^1\gamma^{mn}\theta^2$
$\theta^1\gamma^{rs}\theta^2$ may always be rewritten in terms
of the $SO(9)$ generators $\theta^{1,2}\gamma^{mn}\theta^{1,2}$
or three index bilinears
$\theta^{1,2}\gamma^{mnp}\theta^{1,2}$ via a Fierz rearrangement,
of which
only the latter cause any difficulty.

Clearly, what is needed then is the action of three index operators
$\theta\gamma^{mnp}\theta$ (dropping the labels 1,2) on states
$|-\rangle^{\cal M}$. The result is easily obtained by making the
most general
ansatz and fixing the coefficients via the explicit representation
given in~\cite{pw}, we find
\bea
\theta\gamma^{mnp}\theta|-\rangle^{tu}&=&
-i24\sqrt 3 (\delta^{mt}\,|-\rangle^{npu}
-\ft{1}{9}\delta^{tu}\,|-\rangle^{mnp})\,\, ,\label{alg1}\\
\theta\gamma^{mnp}\theta|-\rangle^{tuv}&=&
i24\sqrt 3\, \delta^{mt}\delta^{nu}\,|-\rangle^{pv}+
\ft{2i}{3}\epsilon^{mnptuvwyz}
|-\rangle^{wyz}\, .\label{alg2}
\eea
On the right hand side of~\eqn{alg1} and~\eqn{alg2} one must
(anti)symmetrize (with unit weight) over all indices according to
the symmetry  properties of the left hand sides of these equations.

\subsection{Results}

Given the algebra~\eqn{alg1} and~\eqn{alg2} along with the
potential~\eqn{POT},
only a moderate amount of computer algebra~\cite{Jos}
 is now required to obtain the Matrix
Theory one-loop three form--three form eikonal scattering amplitude.
Our result consists of 103 terms and is given by (normalising the
$v^4$ term
to unity)
\bea
{\cal A}\!\!\!\!\!&\!\!\!\!\!&\,=\,\frac{1}{q^2}\,\Big\{\!
\,\ft12 v^4     \,  C_{14} C_{23}
+ 3v^2  \Big(       C_{14}(v,q) C_{23} - C_{14}(q,v) C_{23} \Big)
\nn\\&&
\!\!
+v^2   \Big(  \ft12  \ft34 C_{12}(q,q) C_{34} + \ft12 \ft34 C_{12}  
C_{34}(q,q)
                    - \ft32 C_{14}(q,q) C_{23} -       \ft34
C_{13}(q,q) C_{24}
\nn\\&&
          + \ft12\ft94 C_{14}(q,m_1) C_{23}(q,m_1)
          + \ft12\ft94 C_{14}(m_1,q) C_{23}(m_1,q)
          -      \ft94 C_{14}(q,m_1) C_{23}(m_1,q)
\nn\\&&
          + \ft12\ft94 C_{13}(q,m_1) C_{24}(q,m_1)
          + \ft12\ft94 C_{13}(m_1,q) C_{24}(m_1,q)
          -      \ft94 C_{12}(q,m_1) C_{34}(q,m_1)
\nn\\&&
          + \ft12\ft92 C_{12}(q,m_1,q,m_2) C_{34}(m_1,m_2)
          + \ft12\ft92 C_{12}(m_1,m_2) C_{34}(q,m_1,q,m_2)          
\nn\\&&
         - \ft92  C_{13}(q,m_1,q,m_2)C_{24}(m_1,m_2)\Big)
\nn\\&&
  -\ft12 \ft92 C_{12}(v,v) C_{34}(q,q)
  + \ft92 C_{12}(v,q) C_{34}(v,q)
  +\ft12 \ft92 C_{14}(v,q) C_{23}(v,q)
\nn\\&&
  - \ft92  C_{14}(v,q) C_{23}(q,v)
  +\ft12\ft92 C_{14}(q,v) C_{23}(q,v)
  + \ft92C_{13}(v,v)  C_{24}(q,q)
\nn\\&&
  -\ft12\ft92C_{13}(v,q) C_{24}(v,q)
  - \ft12\ft92 C_{13}(q,v) C_{24}(q,v)
  - \ft12\ft92 C_{34}(v,v) C_{12}(q,q)
\nn\\&&
    +\ft12  3 C_{12}(v,q,v,q) C_{34}
     +\ft12 3 C_{34}(v,q,v,q) C_{12}
\nn\\&&
    -6 C_{14}(v,q,v,q) C_{23}
    -3 C_{13}(v,q,v,q) C_{24}
\nn\\&&
     -\ft12 9  C_{12}(m_1,m_2) C_3(v,q,m_1) C_4(v,q,m_2)
     -\ft12 9 C_{34}(m_1,m_2) C_1(v,q,m_1) C_2(v,q,m_2)
\nn\\&&
      +9 C_{24}(m_1,m_2) C_1(v,q,m_1) C_3(v,q,m_2)
\nn\\&&
 - 9 C_{12}(v,q,v,m_1) C_{34}(q,m_1)
 + 9 C_{12}(v,q,q,m_1) C_{34}(v,m_1)
\nn\\&&
 + 9 C_{13}(v,q,v,m_1) C_{24}(q,m_1)
 - 9 C_{13}(v,q,q,m_1) C_{24}(v,m_1)
\nn\\&&
 - 9 C_{13}(m_1,v) C_{24}(q,m_1,v,q)
 + 9 C_{13}(m_1,q) C_{24}(v,m_1,v,q)
\nn\\&&
 + 9 C_{12}(m_1,v) C_{34}(q,m_1,v,q)
 - 9 C_{12}(m_1,q) C_{34}(v,m_1,v,q)
\nn\\&&
 +  \ft12 18 C_{12}(v,m_1,v,m_2) C_{34}(q,m_1,q,m_2)
 +  \ft12 18C_{34}(v,m_1,v,m_2) C_{12}(q,m_1,q,m_2)
\nn\\&&
 - 18  C_{12}(v,m_1,q,m_2) C_{34}(v,m_1,q,m_2)
\nn\\&&
       +\ft94 C_{12}(v,q) C_{34}(q,q)
       -\ft94 C_{14}(v,q) C_{23}(q,q)
       +\ft94 C_{14}(q,v) C_{23}(q,q)
\nn\\&&
       +\ft94 C_{13}(v,q) C_{24}(q,q)
       -\ft94 C_{13}(q,v) C_{24}(q,q)
       -\ft94 C_{34}(v,q) C_{12}(q,q)
\nn\\&&
    +\ft92 C_{12}(v,q,q,m_1) C_{34}(q,m_1)
    -\ft92 C_{14}(v,q,q,m_1) C_{23}(q,m_1)
\nn\\&&
    +\ft92 C_{14}(v,q,q,m_1) C_{23}(m_1,q)
    +\ft92 C_{13}(v,q,q,m_1) C_{24}(q,m_1)
\nn\\&&
 -\ft92 C_{24}(q,m_1,v,q) C_{13}(m_1,q)
 -\ft92 C_{14}(q,m_1,v,q) C_{23}(q,m_1)
\nn\\&&
 +\ft92 C_{14}(q,m_1,v,q) C_{23}(m_1,q)
 -\ft92 C_{34}(q,m_1,v,q) C_{12}(m_1,q)
\nn\\&&
       -9 C_{12}(v,m_1,q,m_2) C_{34}(q,m_1,q,m_2)
       +9 C_{34}(v,m_1,q,m_2) C_{12}(q,m_1,q,m_2)
\nn\\&&
          - \ft12\ft98 C_{12}(q,q) C_{34}(q,q)
          - \ft12\ft98 C_{14}(q,q) C_{23}(q,q)
          - \ft12\ft98 C_{13}(q,q) C_{24}(q,q)
\nn\\&&
          + \ft12\ft92 C_{12}(q,m_1,q,m_2) C_{34}(q,m_1,q,m_2)
          + \ft12\ft92 C_{12}(q,m_1,q,m_2) C_{34}(q,m_2,q,m_1)
\nn\\&&
          + \ft12\ft92 C_{13}(q,m_1,q,m_2) C_{24}(q,m_2,q,m_1)
\nn\\&&
\,+\,\Big[\,C_1\,\longleftrightarrow\, C_2\, ,\,C_3\,
\longleftrightarrow \,C_4\,\Big]
\Big\}
\label{result}
\eea
where we have introduced the notation
in which an $n$-index tensor
$T_{m_1\ldots m_n}$ written as a function of a vector
$q_m$ denotes
$T(q,m_{2},\ldots,m_n)\equiv T_{m_1m_2\ldots m_n}
q_{m_1}$. Moreover the tensors
$C_{ij}$ ($i,j=1,\ldots,4$)
denote the contraction of polarisation tensors,
for example $C_{14}(q,v)=C_1^{m_1m_2m_3}C_4^{m_3m_2n_1}q^{m_1}
v^{\n_1}$ (where the indices are contracted between the two tensors in
the order indicated, another example is
$C_{12}=C_1^{m_1m_2m_3}C_2^{m_3m_2m_1}$).
In the next section we shall see that this result\footnote{We note,  
in passing, that in obtaining
\eqn{result}, products of as many as four nine dimensional 
Levi-Civita symbols
were encountered and expanded in Kronecker deltas.
For convenience, we have also put an explicit $\ft12$ in front of  
terms mapped to themselves under the replacement
$[\,1\,\longleftrightarrow\, 2\, ,\,3\,
\longleftrightarrow \,4\,]$.
} yields perfect agreement
with SUGRA Feynman graphs.

%%%%%%%%%%%%%%%%%%%%%%%%%%%%%%%%%%%%%%%%%%%%%%%%%%%%%%%%%%%%%%%%%%%%%

\setcounter{equation}{0}
\section{The Supergravity Computation.}\label{SUGARandSPICE}
\label{SUGRA}

The bosonic sector of 11-dimensional supergravity \cite{Cremmer} is  
given by
\bea
\lefteqn{{\cal L}= -\ft{1}{2\kappa^2} \sqrt{-g}\, R -\ft 18 \sqrt{-g}\,
(F_{MNPQ})^2}
\nn\\&& -\ft{\sqrt{3}}{12^3\kappa}\varepsilon^{M_1\ldots M_{11}}
F_{M_1M_2M_3M_4}\, F_{M_5 M_6M_7M_8}\, C_{M_9M_{10}M_{11}}
\label{sugra}
\eea
where $F_{MNPQ}=4\partial_{[M}C_{NPQ]}$
and $g={\rm det}\,g_{MN}$.
Perturbative
quantum gravity may be studied by considering small fluctuations $h_{MN}$
from the flat metric $\eta_{MN}$
\be
g_{MN}= \eta_{MN}+ \kappa\, h_{MN}
\ee
where $\kappa$ is the 11-dimensional gravitational coupling
constant. {}From now on we raise and lower indices with the flat metric
$\eta_{MN}$. Propagators are obtained in the usual fashion. For the  
graviton
we employ the harmonic
(de Donder) gauge $\partial_N h^N{}_M-(1/2)\partial_M h^N{}_N=0$
with propagator
(in $d$ dimensions)
\be
\langle
h_{MN}(k)h_{PQ}(-k)\rangle=\frac{-4i}{k^2}\Big(\eta_{(M|P|}\eta_{N)Q}
-\frac{1}{d-2}\eta_{MN}\eta_{PQ}\Big)\, .
\ee
For the antisymmetric tensor, the gauge fixing function
$\partial_M C^M{}_{NP}$
in the weighted gauge ${\cal L}_{\rm fix}=-\frac{3}{2}(\partial^M
C_{MNP})^2$
yields the Feynman propagator
\be
\langle C_{M_1M_2M_3}(k)\, C^{N_1N_2N_3}(-k)\rangle
=\frac{-i}{k^2}\, \d_{[M_1}^{N_1}\, \d_{M_2}^{N_2}\,
\d_{M_3]}^{N_3}.
\label{Cpropagator}
\ee
The relevant vertices are easily read off from~\eqn{sugra}.
In particular, note that $h_{MN}$ couples in the usual way to the
three form stress-energy tensor.

At tree level, the only graphs contributing to four point,
three form scattering are the single graviton and
three form exchange diagrams.
These are easily computed
and in the $t=-2p^1_M p_4^M$ channel, which, as we shall see dominates
eikonal physics (the $s$ and $u$ channels follow anyway by
Bose symmetry) one finds
\bea
{\cal A}_{\bf 84}&=&\frac{-i\kappa^2}{R}\frac{1}{t}\Big\{
\frac{3}{18^2}(\e F^1 F^4)^{MNP}(\e F^2 F^3)_{MNP} \nonumber \\
&&\quad
-32\Big[\Big(\frac{d+16}{d-2}\Big)(F^2\cdot F^3)(F^1\cdot F^4)-
32(F^2\cdot F^3)_{(AB)}(F^1\cdot F^4)^{AB}
\Big]\Big\}\nn\\&& \label{84}
\eea
The factor $R$ in (\ref{84}) represents the radius of the  
compactified light-like
circle,
$F^i_{MNPQ}=p^i_{[M}C^i_{NPQ]}$ ($i=1,\ldots,4$) is the curl of  
the eleven dimensional polarisation tensor $C^i_{MNP}$
and we denote $F^i\cdot F^j=F^i_{MNOP}F_j^{MNOP}$,
$(F^i\cdot F^j)_{AB}=F^i_{AMNP}F^j_{B}{}^{MNP}$
and $(\e F^i
F^j)^{MNP}=\e^{MNPM_1\ldots M_4N_1\ldots N_4}F^i_{M_1\ldots M_4}
F^j_{N_1\cdots N_4}$. Of course, one must put $D=11$ in this formula.
The momenta and polarisations satisfy
the mass shell and gauge conditions $p^i_Mp_i^M=0$ and $p_i^M
C^i_{MNP}=0$,
respectively.

Since we are not aware of any other occurrences of the result~\eqn{84}
in the literature, for added certainty we considered the
dimensional reduction of
the amplitude (\ref{84}) to ten dimensions setting $p^i_{11}=0$. In  
this case, the
three form gauge field, that transforms as an ${\bf 84}$ with
respect to the
little group $SO(9)$, splits into ${\bf 56}$ and ${\bf 28}$
representations of $SO(8)$, that is the little group in ten dimensions.
Then, we checked the validity of~\eqn{84} by
computing the four-point function of the Ramond-Ramond three form
(${\bf 56}$)
and of the antisymmetric tensor field (${\bf 28}$) in type IIA
string theory.
By taking then the low-energy $\alpha^{\prime}\rightarrow 0$ limit  
of the IIA
amplitudes we obtained precisely the dimensional reduction of
(\ref{84})
to ten dimensions.

\subsection{Light Cone Supergravity Amplitudes.}
\label{sugralight}

In order to make a comparison with the Matrix Theory results
we need to rewrite our supergravity $t$ channel amplitude
in terms of physical transverse nine dimensional degrees of freedom.
Since we are considering the $N=2$ Discrete Light Cone
Quantisation (DLCQ) formulation~\cite{suss}
of the theory, we work in
light cone coordinates and specialise to the case of vanishing
$p^-$ momentum exchange.
Define, therefore \footnote{Notice the conventions used here are slightly
different from those used in section three.}
\be
p^{\pm}=p_{\mp}=\frac{p^{10}\pm p^0}{\sqrt{2}}
\ee
so that on-shell momenta satisfy $0=p^Mp_M=2p^+ p^-+p_m p_m$ where
nine dimensional indices $m,n,\ldots$ are contracted
with a Kronecker delta. The gauge condition
$p^MC_{MNP}=0$ may then be solved in terms of 84 physical polarisations
$C_{mnp}$ via
\be
C_{+-m}=0=C_{+mn}\, , \,
C_{-mn}=-\frac{1}{p^-}p_{r}C_{rmn}\, .\label{physical}
\ee
where we have used the residual gauge freedom to set
$C_{+mn}=0$. From now on, we measure momenta in units of the compactified
radius $R$ so that $p^-=1$ and $p^+=-\frac{1}{2}p^m p^m$.

The $N=2$ DLCQ Matrix Theory describes $1+2\longrightarrow 4+3$
scattering in the case of vanishing $p^-$ momentum exchange so we
may therefore write the incoming and outgoing momenta as
 \bea
p_M^1=(-\ft12\,(v_m-q_m/2)^2\, ,\, 1\, ,\, v_m-q_m/2)
&\!\!\!\!\!\!\!\!\!&
p_M^2=(-\ft12\,(v_m-q_m/2)^2\, ,\, 1\, ,\,- v_m+q_m/2) \nonumber\\
 p_M^4=(-\ft12\,(v_m+q_m/2)^2\, ,\, 1\, ,\, v_m+q_m/2) &\!\!\!\!
\!\!\!\!\!\!&
p_M^3=(-\ft12\, (v_m+q_m/2)^2\, ,\, 1\, ,\,-v_m-q_m/2)\nn\\&&
\label{kinematics}
\eea
where we have set the nine dimensional centre of mass momentum to zero by
transverse Galilean invariance.
Importantly, note that conservation of $p^+$ momentum implies $v_m  
q_m=0$.
The Mandelstam variables in this parametrisation read
\be
t=q_m^2=-2p^1_Mp_4^M\, ,\quad s=4v_m^2+q_m^2=-2p^1_Mp_2^M \,
\quad u=4v_m^2=-2p_M^1p_2^M=s-t\, .\label{mandelstam}
\ee
Eikonal scattering, for which the scattering angle
${\rm Cos}^{-1}([v^2-q^2/4]/[v^2+q^2/4])\sim \sqrt{q^2/v^2}$ is small
takes place for small $q^2=t$. Therefore we must study the
amplitude~\eqn{84} in the small $t$ limit which is dominated by the  
$t$-pole.
In fact, we argued above (see section three) that our
D0-brane computation was not reliable for contact terms
proportional to $q^2$,
and therefore in this work we only study those terms
which do not cancel
the $t$-pole
(one may therefore disregard all terms with an $s $ or $u$ pole).
We will further discuss this restriction in the conclusion.

It only remains now to state our main result, namely
that substituting equations~\eqn{physical}, \eqn{kinematics}
and~\eqn{mandelstam} into the amplitude~\eqn{84} and
neglecting terms cancelling the $t$-pole,
one reproduces precisely the 103 terms of the Matrix Theory
amplitude~\eqn{result}.

%%%%%%%%%%%%%%%%%%%%%%%%%%%%%%%%%%%%%%%%%%%%%%%%%%%%%%%%%%%%
\newpage
\setcounter{equation}{0}
\section{Conclusions}
\noindent

In this paper we have computed  a four-point scattering amplitude
and shown, analogously to the graviton case analysed in  
\cite{psw}, that Matrix Theory also reproduces correctly the  
tensorial structures
of the $D=11$ SUGRA three form couplings.
Both the present and our  previous matrix computation have been
performed by considering only the leading one-loop D0-brane effective
potential that is protected by supersymmetry from quantum  
corrections\cite{pss1}. Therefore the agreement found might have  
been expected as is
the case for most of the one-loop
phase shift calculations appearing in the literature.
Furthermore, in principle, it should be possible to fix
the structure and coefficients of equation (\ref{POT})
by supersymmetry alone \cite{pss1}.
However, without a formalism in which amplitudes can be derived
from the Matrix potential (\ref{POT}), it would have remained
unclear how such a potential could be compared with
the tensorial structure of the SUGRA amplitude.
Clearly our work resolves this issue and renders
the precise relationship between
supersymmetry in each of these models more transparent.
We also remark that our formalism might provide a route to
establishing eleven dimensional Lorentz invariance of Matrix Theory.

Another issue deserving comment is the choice of background
about which one perturbatively expands $D=11$ SUGRA
in order to compare with Matrix Theory.
For the case of finite (and actually small) $N$, it is clearly natural
to expand about a flat background and
ignore the geometry induced by the D-particles themselves
(since in that case these states appear as fundamental Kaluza-Klein  
states,
i.e., excitations of the flat vacuum), in contrast to the large $N$ case
when they can be represented as classical sources, modifying then the
background geometry.

Throughout this paper we have considered $t$ channel amplitudes only.
Since we consider quantum asymptotic states which can describe in and
outgoing particles scattering at any angle,
this restriction can in principle be relaxed, although
it is not completely clear how to compute the transition element
(\ref{S})
(or equivalently the corresponding path-integral (\ref{PIBRS}))
in this case
\footnote{In the context of classical gravity source-probe approach,
recoil effects have been recently taken into account in~\cite{oy}.}.
Furthermore, our current work is also limited to the case of
vanishing $p^-$ momentum exchange; although of considerable interest,
interactions involving $p^-$ exchange are clearly not visible in
our perturbative
Matrix Theory framework.

Finally, an important line of development could be to use our
technique to
analyse higher order Matrix Theory scattering amplitudes that would
correspond to, say, one-loop effects in supergravity. This kind of  
comparison
will put Matrix Theory to a much more stringent test and will be crucial
in trying to understand the range of validity of the theory itself.

\vspace{1cm}
%%%%%%%%%%%%%%%%%%%%%%%%%%%%%%%%%%%%%%%%%%%%%%%%%%%%%%%%%%%%
\noindent {\large \bf Acknowledgements}

We wish to thank C.A. Scrucca for important discussions and
correspondence. In our computations we have made extensive use of
the computer algebra
system FORM~\cite{Jos}. This work was supported in part by the
Nederlandse Organisatie voor Wetenschappelijk Onderzoek (NWO).

%%%%%%%%%%%%%%%%%%%%%%%%%%%
\vspace{1cm}
\renewcommand{\theequation}{A.\arabic{equation}}
\setcounter{equation}{0}
\par \noindent
  {\Large \bf A. Appendix}
  \par
   \vspace{2mm} %was 5 mm
\noindent
%%%%%%%%%%%%%%%%%%%%%%%%%%%%

\noindent
We employ the following Dirac matrix conventions,
the $32\times 32\,$ $SO(1,9)$ matrices are
\be
\Gamma^0 = \pmatrix {0 & I_{(16)} \cr -I_{(16)} & 0 \cr}, \ \ \
\Gamma^i= \pmatrix {0 & \gamma^i_{(16)} \cr \gamma^i_{(16)} & 0
\cr}, \ \ i=1,...,9
\ee
where the $SO(9)$ Dirac matrices are chosen to be real and symmetric
\be
\gamma^9_{(16)} = \pmatrix {I_{(8)} & 0 \cr 0 & -I_{(8)} \cr}, \ \  
\gamma^i_{(16)}= \pmatrix {0 & \gamma^i_{(8)} \cr
(\gamma^i_{(8)})^{\rm T} & 0 \cr},
\ \ i=1,...,8
\ee
The $SO(8)$ Dirac matrices $\gamma^i_{(8)}$ are the same as those
appearing in
volume one of \cite{gsw}.
The charge conjugation matrix $C$ is identified with $\Gamma^0$ so  
that for a Majorana
spinor $\bar{\Psi}=\Psi^{\rm T}\Gamma^0$. In section three
use has been made of the following
Fierz identities
\be
\bar{\Psi}\Gamma^{\mu\nu\rho}\Psi\;\bar{\Psi}
\Gamma_{\mu\nu}^{\;\;\;\;\sigma}\Psi=0
\ee
which in $SO(9)$ and $SO(8)$ language read
\bea
\hspace{-2cm}
&\!\!&(\theta\gamma^{ijk}_{(16)}\theta)(\theta\gamma^{ijl}_{(16)}\theta)-2
(\theta\gamma^{ik}_{(16)}\theta)(\theta\gamma^{il}_{(16)}\theta)=0   
\label{fi9} \\ \hspace{-2cm}
&\!\!&(\eta\gamma^{ijk}_{(8)}\widetilde{\eta})(\eta\gamma^{ijl}_{(8)}
\widetilde{\eta})
-2(\eta\gamma^{k}_{(8)}\widetilde{\eta})(\eta\gamma^{l}_{(8)}\widetilde{\eta})
-(\eta\gamma^{ik}_{(8)}\eta)(\widetilde{\eta}\gamma^{il}_{(8)}\widetilde{\eta})
-(\eta\gamma^{il}_{(8)}\eta)(\widetilde{\eta}\gamma^{ik}_{(8)}
\widetilde{\eta})=0
\label{fi4}
\eea
with $\Psi=\left(^{\textstyle \theta}_{\textstyle 0}\right)$,
$\theta = (\eta^a, \widetilde \eta^{\dot a})$ and the common   
notation$\gamma^{i_1...i_n}\equiv  
(1/n!)\,\gamma^{i_1}...\gamma^{i_n}\pm{\rm perm.}$
valid for all gamma matrices.

The expressions (\ref{3v1}),(\ref{2v0})
reduce to the form appearing in the right hand sides of
equations~(\ref{2v1}) and (\ref{3v0}), using the relation
\be
q_{\alpha}q_{\beta} \bar{\Psi}\Gamma^{\mu_1\mu_2\alpha}\Psi\;
\bar{\Psi}\Gamma^{\mu_3\mu_4\beta}\Psi \sim
q_{\alpha}q_{\beta}\,
\bar{\Psi}\Gamma^{\alpha}_{\;\;\;\mu\nu}\Psi\;
\bar{\Psi}\Gamma^{\mu_1\mu_2\mu_3\mu_4\beta\mu\nu}\Psi + \ldots
\label{fi5}
\ee
where dots stand for terms proportional to
$\delta^{\mu_i}_{\alpha},\delta^{\mu_i}_{\beta},\delta^{\mu_i}_{\mu_j}$  
that vanish
in (\ref{3v1}),(\ref{2v0}) and  contact terms proportional to
$\delta^{\alpha}_{\beta}$ have been neglected. In order to write
our potential in
the form shown in (\ref{POT}), the $SO(9)$ Fierz identity (A.5)
of \cite{static}
as well as (\ref{fi9}) are needed.
We used moreover the Fierz identity
\be
(\eta\gamma^{[ij}\eta)(\widetilde{\eta}\gamma^{kl]}\widetilde{\eta})
=-\frac{1}{24}
\epsilon^{ijklmnpq}(\eta\gamma^{mn}\eta)(\widetilde{\eta}\gamma^{pq}
\widetilde{\eta})
\label{fi6}
\ee
as well as those involving the $\epsilon$ tensor
\bea
&\!\!&{\bf (A)}\cdot q_{i_1}(\eta\gamma^{i_2}\widetilde{\eta})
(\widetilde{\eta}\gamma^{i_3i_4m}\eta) \sim {\bf (A)}\cdot q_{i_1}\,
[(\eta\gamma^{i_2m}\eta)(\widetilde{\eta}\gamma^{i_3i_4}\widetilde{\eta})-
(\eta\gamma^{i_3i_4}\eta)(\widetilde{\eta}\gamma^{i_2m}\widetilde{\eta})];  
\nonumber \\
&\!\!&{\bf (A)}\cdot q_k (\eta\gamma^{i_1i_2k}\widetilde{\eta})
(\widetilde{\eta}\gamma^{i_3i_4m}\eta) \sim {\bf (A)}\cdot q_{i_4}\,
[(\eta\gamma^{i_1i_2}\eta)(\widetilde{\eta}\gamma^{mi_3}\widetilde{\eta})+
(\eta\gamma^{mi_3}\eta)(\widetilde{\eta}\gamma^{i_1i_2}\widetilde{\eta})];  
\nonumber \\
&\!\!&{\bf (A)}\cdot q_{i_1}(\eta\gamma^{m}\widetilde{\eta})
(\widetilde{\eta}\gamma^{i_2i_3i_4}\eta) \sim {\bf (A)}\cdot q_{i_1}\,
[(\eta\gamma^{i_2m}\eta)(\widetilde{\eta}\gamma^{i_3i_4}\widetilde{\eta})-
(\eta\gamma^{i_3i_4}\eta)(\widetilde{\eta}\gamma^{i_2m}\widetilde{\eta})];  
\nonumber   \\
&\!\!&{\bf (B)}\cdot (\widetilde{\eta}\gamma^{i_2i_3m}\eta)
(\eta\gamma^{ni_4}\eta) \sim {\bf (B)}\cdot
(\eta\gamma^{nl}\eta)(\eta\gamma^{i_2i_3i_4lm}\widetilde{\eta});
\label{fi7} \\
&\!\!& \epsilon^{1ii_1...i_6}\,v_i\,q_{i_1}\,q_m\,q_n
(\eta\gamma^{i_6n}\eta)
(\eta\gamma^{i_2i_3i_4}\widetilde{\eta})
(\widetilde{\eta}\gamma^{i_5m}\widetilde{\eta}) \sim \nonumber \\
&\!\!& \epsilon^{1ii_1...i_6}\,v_i\,q_{i_1}\,q_m\,q_n\,
[(\widetilde{\eta}\gamma^{lm}\widetilde{\eta})
(\eta\gamma^{i_2i_3i_4i_5l}\widetilde{\eta})
+2(\widetilde{\eta}\gamma^{i_5l}\widetilde{\eta})
(\eta\gamma^{i_2i_3i_4lm}\widetilde{\eta})]; \nonumber \\
&\!\!& \epsilon^{i_1...i_8}q_{i_1}q_mq_nq_k(\eta\gamma^{i_2m}\eta)
(\widetilde{\eta}\gamma^{i_3n}\widetilde{\eta})
(\eta\gamma^{i_4i_5k}\widetilde{\eta})(\widetilde{\eta}
\gamma^{i_6i_7i_8}\eta) \sim \nonumber \\
&\!\!& \epsilon^{i_1...i_8}q_{i_1}q_mq_nq_k(\eta\gamma^{i_2m}\eta)
(\widetilde{\eta}\gamma^{i_3n}\widetilde{\eta})
[(\widetilde{\eta}\gamma^{kl}\widetilde{\eta})
(\eta\gamma^{li_4...i_8}\widetilde{\eta})
+(\widetilde{\eta}\gamma^{li_8}\widetilde{\eta})
(\eta\gamma^{i_4...i_7kl}\widetilde{\eta})] \nonumber
\eea
where
\bea
{\bf (A)} &=& \left\{
\begin{array}{l}
 \epsilon^{1ii_1...i_6}\,v_i\,q_m\,q_n\,
(\eta\gamma^{i_5i_6n}\widetilde{\eta}) \nonumber
\medskip\ \\
\epsilon^{i_1...i_8}\,q_m\,q_n\,q_p\,(\eta\gamma^{i_5i_6n}\widetilde{\eta})
(\eta\gamma^{i_7i_8p}\widetilde{\eta})
\end{array}
\right. \\
{\bf (B)} &=& \left\{
\begin{array}{l}
\epsilon^{1ii_1...i_6}\,v_i\,q_{i_1}\,q_n\,q_m\,
(\widetilde{\eta}\gamma^{i_5i_6}\widetilde{\eta})
\medskip\ \\
\epsilon^{i_1...i_8}\,q_{i_1}\,q_p\,q_n\,q_m\,
(\widetilde{\eta}\gamma^{i_5i_6}\widetilde{\eta})
(\eta\gamma^{i_7i_8p}\widetilde{\eta})
\end{array} \nonumber
\right.
\eea
valid, respectively, for the linear in $v$ and static term in the
potential. As explained in section three, in order to fix the
relative factors
between the two contributions appearing in the linear in $v$ and
static effective
potential, there is no need to know the coefficients in
(\ref{fi5}) and (\ref{fi7}),
as it is by far more convenient to use only (\ref{fi6}).

%%%%%%%%%%%%%%%%%%%%%%%%%%%%%%%%%%%%%%%%%%%%%%%%%%%%%%%%

\end{document}